\documentclass[conference]{IEEEtran}
\usepackage{flushend}
\usepackage{float}
\restylefloat{table,tabular}
\usepackage{amsmath,mathrsfs,graphics,graphicx}

\ifCLASSINFOpdf

\else

\fi

\newcommand{\Trans}[1]{\mbox{$\stackrel{#1}{\Longrightarrow}$}}

\newcommand{\proc}{\mathcal{P}\hspace{-1.5pt}{\it roc}}
\newcommand{\const}{\mathcal{C}\hspace{-1.5pt}{\it onst}}
\newcommand{\config}{\mathcal{C}\hspace{-1.5pt}{\it onf}}

\newcommand{\rtcc}{\mbox{{\bf{\footnotesize\textsc{ttcc}}}}}

\newcommand{\nul}{\mathbf{0}}
\newcommand{\tell}{\mathbf{tell}}
\newcommand{\when}{\mathbf{when}}
\newcommand{\doo}{\mathbf{do}}

\newcommand{\local}{\mathbf{local}}
\newcommand{\inn}{\mathbf{in}}

\newcommand{\next}{\mathbf{next}}

\newcommand{\deff}{\stackrel{ \mbox{def} }{=}}
\newcommand{\ima}{\mathbf{IMA}}
\newcommand{\tte}{\mathbf{TTE}}
\newcommand{\avio}{\mathbf{AVIO}}
\newcommand{\maf}{\mathbf{MAF}}
\newcommand{\cf}{\mathbf{CF}}
\newcommand{\lcm}{\mathbf{LCM}}
\newcommand{\true}{\mathbf{true}}

\newcommand{\entail}{\vdash}

\newcommand{\vertex}{\mathcal{V}}
\newcommand{\edge}{\mathcal{E}}
\newcommand{\link}{\mathcal{L}}
\newcommand{\vlink}{\mathcal{V}\hspace{-1.5pt}\mathcal{L}}
\newcommand{\dpath}{\mathcal{D}\hspace{-1.5pt}\mathcal{P}}
\newcommand{\vl}{v\hspace{-0.5pt}l}

\newcommand{\ttsched}{\tilde{\varsigma}}
\newcommand{\wf}{\mathbf{WF}}
\newcommand{\sr}{\mathbf{SR}}
\newcommand{\lt}{\mathbf{LT}}
\newcommand{\fd}{\mathbf{FD}}

\begin{document}
%
\title{A Time-Triggered Constraint-Based Calculus for Avionic Systems
}

\author{
\IEEEauthorblockN{ Sardaouna Hamadou\IEEEauthorrefmark{1}, Abdelouahed Gherbi\IEEEauthorrefmark{2}, John Mullins\IEEEauthorrefmark{1} and Sofiene Beji\IEEEauthorrefmark{1}\\}
\\
\IEEEauthorblockA{\IEEEauthorrefmark{1}Department of Computer and Software Engineering\\
\'{E}cole Polytechnique de Montr\'{e}al\\
Montreal (Quebec), Canada\\
Email: firstname.lastname@polymtl.ca \\}
\\
\IEEEauthorblockA{\IEEEauthorrefmark{2}Department of Software and IT Engineering\\
\'{E}cole de Technologie Sup\'erieure\\
Montreal (Quebec), Canada \\
Email: abdelouahed.gherbi@etsmtl.ca}

}

\maketitle

\begin{abstract}
The Integrated Modular Avionics (\emph{IMA}) architecture and 
the Time-Triggered Ethernet (TTEthernet) network have emerged as the 
key components of a typical architecture model for recent civil aircrafts.  
We propose a real-time constraint-based calculus targeted at the analysis 
of such concepts of avionic embedded systems. We show our framework at work 
on the modelisation of both the (\emph{IMA}) architecture and the
TTEthernet network, illustrating their behavior by the well-known
Flight Management System (\emph{FMS}).
\end{abstract}

\section{Introduction}
\label{sec:intro}

The growing complexity of avionic embedded systems led to the definition 
of a new standard of architecture called \textit{Integrated Modular Avionics} 
(\emph{IMA})~\cite{Arinc653}. This type of architecture is characterized 
essentially by the sharing of distributed computing resources, called modules. 
Sharing these resources requires to guarantee some safety properties such as 
the collision-free. In order to achieve this objective, 
the \textit{Avionic Full Duplex Switched Ethernet} (\emph{AFDX})~\cite{AFDX} 
has been adopted as a networking standard for the avionic systems. 
\emph{IMA} and \emph{AFDX} became the components of a typical architecture model 
for the recent civil aircrafts such as the Boeing $B787$ and the Airbus $A380$. 
However, \emph{AFDX} underuses the physical capacities of the network. 
Recently, the \textit{Time-Triggered Ethernet} (TTEthernet) has emerged 
as a new standard of the avionic network~\cite{TTE2011}. 
This standard enables to achieve a best usage of the network 
and is more deterministic since the schedule is established offline.

Both \emph{IMA} and TTEthernet segregate mixed-criticality components into partitions
for a safer integration. \emph{IMA} enables applications to interact safely by partitioning 
them spatially (memory zones) and temporally (processor schedules) over distributed
Real-Time Operating Systems (RTOS). TTEthernet, on the other hand, allows these 
distributed RTOS to communicate safely with each other by partitioning bandwidth 
into time slots (network schedules).

 Although a considerable effort has recently been devoted for the validation of TTEthernet
(e.g.~\cite{MikucionisLRNSPPH10,BehjatiYNBS11}),
to the best of our knowledge the analysis and validation of TTEthernet usage in
model-based development for the integration on \emph{IMA} has been so far relatively ignored.
The present paper proposes a framework which empowers us to analyze, as well as provide guidelines 
and some mechanism design principles for \emph{IMA} integration through TTEthernet network, 
exploiting notions and techniques from Concurrent Constraint Programming (CCP) formalism.

CCP~\cite{SaraswatRP91,SaraswatCCP93} is a well-established formalism 
for reasoning about concurrent and distributed systems. It is a matured model 
of concurrency with several reasoning techniques (e.g.~\cite{Sangiorgi2011,BoerGM00,AlpuenteGPV08}) 
and implementations (e.g.~\cite{SaraswatJG03,Smolka98,LescaylleV09}).
It is adopted in a wide spectrum of domains and applications such as \textit{biological phenomena}, 
\textit{reactive systems} and \textit{physical systems}. 
It enjoys a dual view of processes as agents interacting with one another  
and as logical formulas allowing to benefit from both the well-established process calculi 
and logic formalisms. CCP is a powerful way to define complex synchronization schemes 
in concurrent and distributed settings parametric in a constraint system. 
This  provides  a very flexible way to tailor data structures to specific domains 
and applications. We refer the reader to~\cite{OlarteRV13} for a recent survey on ccp-based models.

\begin{figure*}[!t]
\centering
  \includegraphics[width=0.7\linewidth]{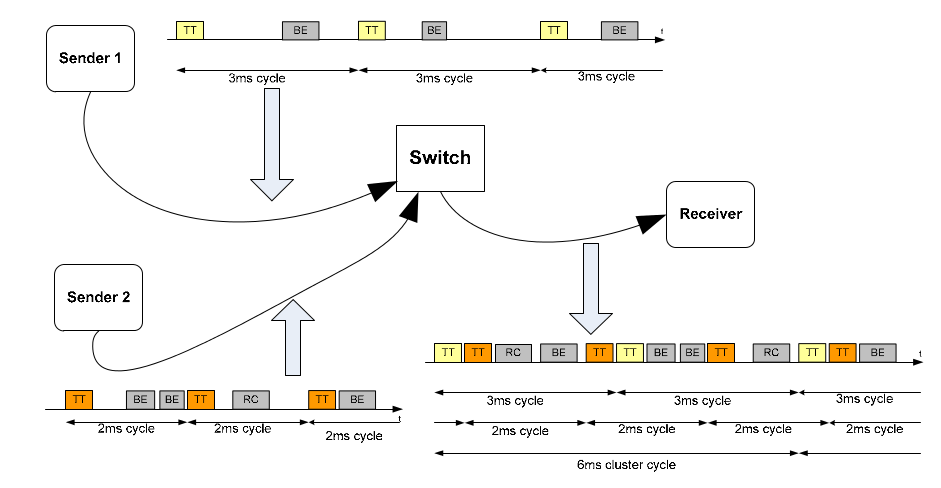} \\
  \caption{\cite{TTE2011} Example of a Time-Triggered Ethernet network}
  \label{fig:TTE}
\end{figure*}

Drawing on earlier work on timed ccp-based formalisms~\cite{SaraswatJG94,NielsenPV02}, 
in the present paper we introduce a calculus to provide a formal 
basis for the analysis of time-triggered architectures in 
avionic embedded systems. Like its predecessors, our calculus 
is built around a small number of primitive operators or combinators parametric
in a \textit{constraint system}.
It extends the Timed Concurrent Constraint Programming (tcc)~\cite{SaraswatJG94} 
in order to define infinite periodic behaviors specific to \emph{IMA} and TTEthernet and
to comply with the requirements of time triggered processes. We then demonstrate 
the pertinence of this calculus by an elegant modeling of the concepts related to 
the \emph{IMA} and TTEthernet architectures. Finally, we illustrate these concepts 
by modeling a subsystem of the well-known flight management system.

To our knowledge, the present paper is the first to provide
a comprehensive framework for the behavioral analysis  for
\emph{IMA} systems deployed throughout a TTEthernet network.
Previous process algebraic models do not deal with both 
\emph{IMA} and TTEthernet~\cite{BremondLG93,SokolskyLC09,PhilippouLS12}, 
or only accounted for the \emph{IMA} concept without
providing a comprehensive set of reasoning techniques 
for the verification of the requirements of avionic systems.
Similarly, existing formal calculi such as the Network Calculus~\cite{LeBoudec01:NC}
and the Real-Time Calculus~\cite{Samarjit00RTC} fall short
of accounting for the time triggered architecture while maintaining a good
accuracy in specifying the system designs~\cite{LiSF10}.

 The rest of the paper is organized as follows: 
 in \S\ref{sec:prelim} we fix some basic notations, briefly revise the concepts 
 of \emph{IMA} and TTEthernet architectures, and introduce the flight management system; 
 in \S\ref{sec:rtcc} we present our conceptual framework; 
 \S\ref{sec:ima}, \S\ref{sec:tte} and \S\ref{sec:avio} deliver our core technical contribution:
 the modelisation of \emph{IMA} systems deployed throughout a TTEthernet network; 
 \S\ref{sec:concl} contains our concluding remarks.

\section{Preliminaries}
\label{sec:prelim}

This section briefly revises the concepts of \emph{IMA} and TTEthernet architectures which underpin the work in this paper.
It also introduces the flight management system, a leading example used to illustrate our conceptual framework.

\subsection{Integrated Modular Avionics}
\label{sec:IMA-concept}

The main idea underlying the concept of \emph{IMA} architecture~\cite{al2012strictly} 
is the sharing of resources between some functions while preventing any interference between them. 
Resource sharing reduces the cost of large volume of wiring and equipment while the non interference 
guarantee is required for safety reasons.

 The \emph{IMA} architecture is a modular real-time architecture for avionic systems defined in ARINC653 \cite{Arinc653}. 
 Each functionality of the system is implemented by one or a set of functions distributed across different modules. 
 A module represents a processor where many functions can be executed. Functions deployed on the same module 
 may have different criticality levels. For safety reasons, the functions  must be strictly isolated using partitions. 
 The partitioning of these functions is two dimensional: spatial partitioning and temporal partitioning. 
 The spatial partitioning is implemented by assigning statically all the resources for the partition 
 being executed in a module and no other partition can have the access to the same resources at the same time. 
 The temporal partitioning is rather implemented by allocating a periodic time window dedicated 
 for the execution of each partition.

\subsection{Time-Triggered Ethernet }
\label{sec:TTE-archit}

Ethernet uses an {\it event-triggered transfer} principle where an end system can access the network 
at arbitrary points in time. Service to the end systems is on a first come, first serve basis which, 
unfortunately, can substantially increase {\it transmission delay} and  {\it jitter} when 
several end systems need to communicate over the same shared medium. Time-Triggered Ethernet~\cite{TTE2011} 
(TTEthernet) specifies time-triggered services that are added to the standards for Ethernet
established in IEEE STD 802.3-2005. In contrast to event-triggered transfer principle, the 
{\it time-triggered transfer} principle uses a network-wide synchronized time base to 
coordinate between end systems, which limits latency and jitter. As depicted in Figure~\ref{fig:TTE}, 
TTEthernet enables time-triggered and event-triggered communication, as well as integrated 
time-triggered/event-triggered communication on the same physical network.  TTEthernet limits 
latency and jitter for time-triggered (TT) traffic, limits latency for rate-constrained (RC) traffic, 
while simultaneously supporting the best-effort (BE) traffic service of IEEE 802.3 Ethernet. 
This allows application of Ethernet as a unified networking infrastructure. In this paper, however,
we shall consider only time-triggered traffic\footnote{We leave the integration of event-triggered 
communication for future work.} (TT).  

\begin{figure}[!tb]
\setlength{\unitlength}{1cm}
\centerline{
\includegraphics[width=8cm]{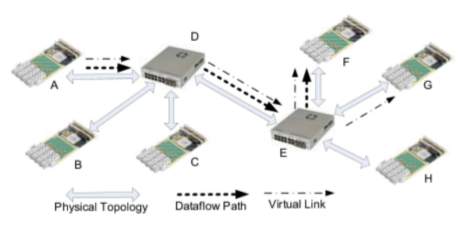}
}
\caption{\cite{steiner2010evaluation} A TTEthernet with six end systems and two switches connected in multi-hop}
\label{fig:TTEtopology}
\end{figure}

As depicted in Figure~\ref{fig:TTEtopology}, the physical topology of a TTEthernet network is 
a graph $\mathbf{G}(\vertex, \edge)$, where end systems and switches are vertices 
$\vertex$ and the physical links connecting  vertices are edges $\edge$. Each physical link connecting
two vertices defines {\em two directed "dataflow links"}. The set of dataflow links is denoted $\link$.
We denote by $[v_1, v_2]$ the dataflow link from vertex $v_1$ to vertex $v_2$ and by 
$$p=[[v_1, v_2], [v_2, v_3], \cdots [v_{m-2}, v_{m-1}],[v_{m-1}, v_m]]$$ the dataflow path connecting one 
end system (the sender) $v_1$ to exactly one other end system (the receiver) $v_m$. 
In Figure~\ref{fig:TTEtopology}, a path from A to F is depicted by the dotted line. In accordance with
the Ethernet convention, information between the sender and receiver is communicated in form of messages $f_i$
called {\em frames}. $\mathcal{F}$ denotes the set of all frames. Frames may be delivered from a sender 
to multiple receivers where the individual dataflow paths between the sender and each single receiver 
together form a ''virtual link``. Hence, a virtual link $\vl$ is the union of the dataflow paths that link
the sender to each receiver. An example of a virtual link from the sender A to the receiver F and G 
is shown in Figure~\ref{fig:TTEtopology}. We denote by $\dpath$ (resp. $\vlink$) the set of 
dataflow paths (resp. virtual links).

\subsection{Flight Management System}
\label{sec:fms}

\begin{figure}[!t]
 \setlength{\unitlength}{1cm}
 \centerline{
 \includegraphics[width=8cm]{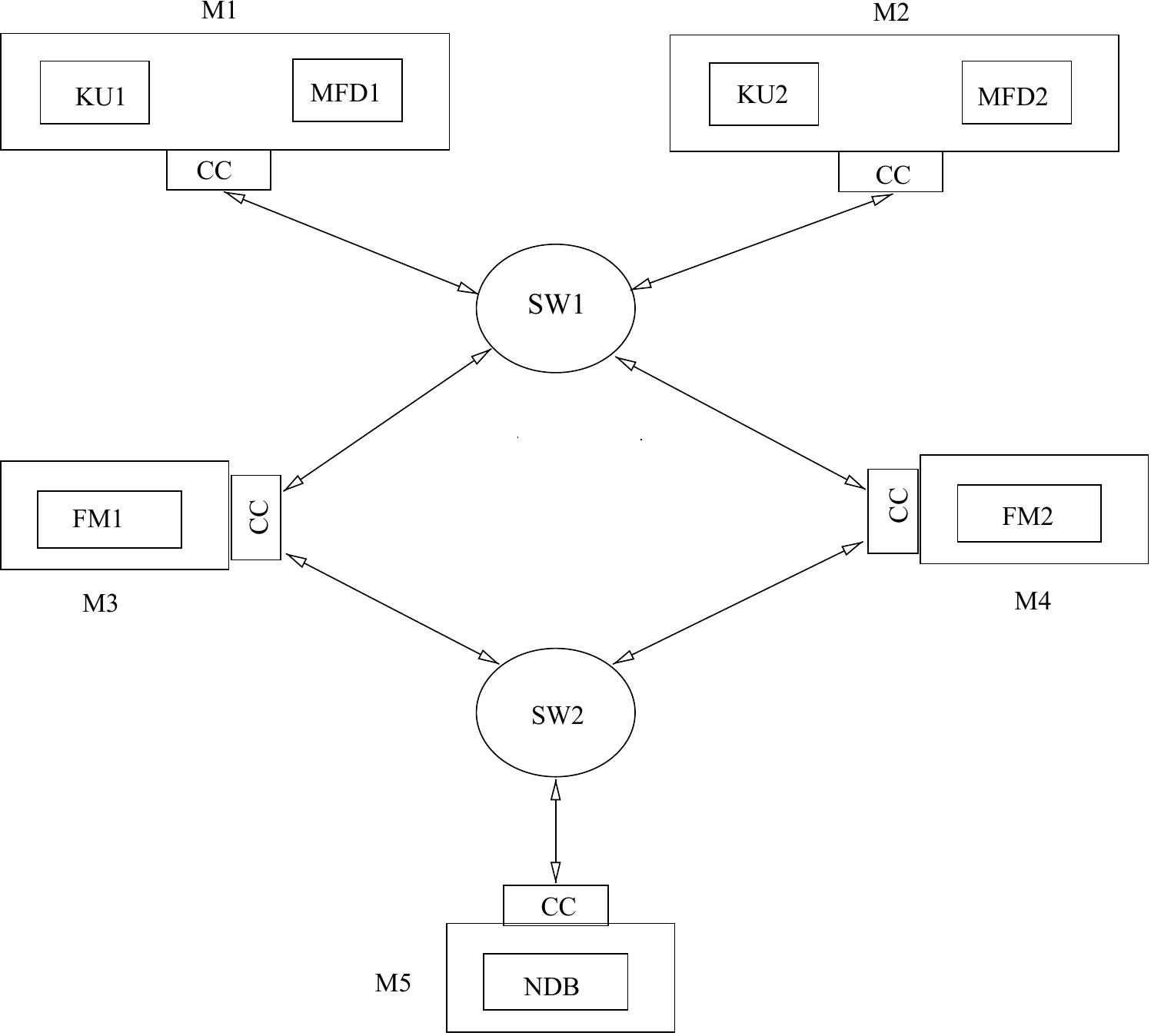}
 }
 \caption{Flight Management System}
 \label{fig:FMS}
 \end{figure}

Now, we briefly recall a subsystem of the Flight Management System drawn from\cite{lauer2012methode}. 
It controls the display of static navigation information in the cockpit screens and it is illustrated by 
Figure~\ref{fig:FMS}.

\begin{figure}[!h]
\setlength{\unitlength}{1cm}
\centerline{
\includegraphics[width=9cm]{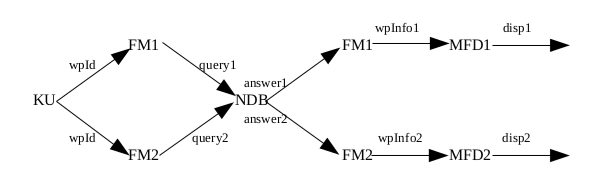}
}
\caption{FMS dataflow}
\label{fig:FMSdataFlow}
\end{figure}

Two partitions  $\mbox{\emph{KU}}_i$ ($i=1;2$), the \emph{Keyboard and cursor Unit}, 
handle the waypoint information requests of the pilot and the co-pilot. 
Each request is transmitted to  $\mbox{\emph{FM}}_1$ and $\mbox{\emph{FM}}_2$, 
the two \emph{Flight Manager} partitions. Each \emph{FM} requests separately \emph{NDB}, 
the \emph{Navigation DataBase}, via an unicast communication to retrieve the waypoint information. 
Each $\mbox{\emph{FM}}_i$ ($i=1;2$) transmits the result to its associated $\mbox{\emph{MFD}}_i$, 
the \emph{Multi Functional Display} partition, which displays the information on the corresponding screen. 
The dataflow inside this subsystem is summarized by Figure~\ref{fig:FMSdataFlow}.

%

%
%
%
 \begin{table*}[!tb]
 	\begin{center}
 \begin{tabular}{|ccc|}
 \hline
 &&\\
 (R-Tell) $\frac{}{\langle \tell(c), d \rangle \longrightarrow \langle \nul, d \wedge c \rangle}$
 & \  
 	(R-Ask-1) $\frac{ d \entail c }
 	         {\langle \when \ c \ \doo \ P , d \rangle \longrightarrow \langle P, d \rangle }$ 
 & \ (R-Ask-2) $\frac{ d \not\entail c }{\langle \when \ c \ \doo \ P , d \rangle \longrightarrow \langle \nul, d \rangle }$ \\ [6mm]	
	\multicolumn{3}{|c|}
	{
	$\begin{array}{ll}
 	\mbox{(R-Par)} \ \frac{\langle P, d \rangle \longrightarrow \langle P', d_p' \rangle \ \ \langle Q, d \rangle \longrightarrow \langle Q', d_q' \rangle}
        {\langle P \; \Vert \; Q, d \rangle \longrightarrow \langle P' \; \Vert \; Q', d_p' \wedge d_q' \rangle}
	 & \
	\mbox{(R-Loc)} \ \frac{ \langle P, c \wedge \exists x d \rangle \longrightarrow \langle P', c'\wedge \exists x d \rangle }
 	{\langle (\local \ x; c) \ \inn \ P , d \rangle \longrightarrow \langle (\local \ x; c') \ \inn \ P', d \wedge \exists x c' \rangle} \\ [6mm]	
	\mbox{(R-Per)} \ \frac{}
 	            {\langle !_{T} P, d \rangle \longrightarrow \langle P \; \Vert \; \next^T (!_{T} P), d \rangle }
	& \
	\mbox{(R-Def)} \ \frac{A(\tilde{x}) \deff P \ \ \langle P[\tilde{v}/\tilde{x}] , d \rangle \longrightarrow \langle P', d' \rangle}
 	            {\langle A(\tilde{v}), d \rangle \longrightarrow \langle P', d' \rangle }
	\end{array}$
		}\\
 &&\\
 \hline 
\hline 	
&&\\
 \multicolumn{3}{|c|}
 	{(R-Obs) $\frac{\langle P, c \rangle \longrightarrow^\ast   \langle Q, d  \rangle \not\longrightarrow \ \  F(Q)=R}
 	            {P \ \Trans{(c,d)} \ R }$}\\
 &&\\
 \hline

 \end{tabular}
 \end{center}
 \caption{Internal transition rules $\longrightarrow$ (upper part) and the observable transition rule $\Longrightarrow$ (lower part).}
 \label{tab:semantics}
 \end{table*}
\section{The ttcc Process Calculus}
\label{sec:rtcc}

This section describes the syntax and the operational semantics of the Time-Triggered  Concurrent Constraint Programming  (ttcc). This calculus
extends the Timed Concurrent Constraint Programming (tcc)~\cite{SaraswatJG94} both syntactically and
semantically. On the syntactic level, we add a new operator to define infinite periodic behaviors 
specific to IMA and TTEthernet. On the semantics level, we extend tcc model to comply with the 
requirements of time triggered processes.
We start by recalling a fundamental notion in ccp-based calculi: {\em constraint system}.

\subsection{Constraint System}
\label{sec:const}

The $\rtcc$ model is parametric in a constraint system specifying the structure and interdependencies of
information that processes can ask of and add to a {\em  central shared store}. A constraint system provides a signature
(a set of constants, functions and predicates symbols) from which constraints\footnote{We remind the reader that a 
constraint represents a piece of information upon which processes may act.} can be constructed as well as 
an entailment relation (a first order logic over the signature), denoted $\entail$, which specifies the interdependancies between these constraints. 
Formally, a constraint system is a pair $(\Sigma, \Delta)$ where $\Sigma$ is a signature and $\Delta$ is a first order theory
 over $\Sigma$. Constraints are first-order formulas over $\mathcal{L}$, the underlying first-order language under $\Sigma$.
%
%
We shall denote by $\const$ the set of constraints 
in the underlying constraint system with typical elements $c, d, \ldots$. 
Given two constraints (i.e. two pieces of information) $c$ and $d$, we say that $c$ entails $d$,
and write $c \entail d$, if and only if $c \Rightarrow d$ is true in all models of $\Delta$. 
In other words, $d$ can be deduced from $c$.

\subsection{Process Syntax}
\label{sec:syntax}

As shown in Figure~\ref{fig:ccp}, a common store is used as communication medium by $\rtcc$ processes to post and read 
constraints (viz. partial information). We use $\proc$ to denote the set of all
$\rtcc$ processes, with typical elements $P$, $Q$, $\ldots$. 
They are built from the following primitive operators: 
\begin{figure}[!tb]
\setlength{\unitlength}{1cm}
\centerline{
\includegraphics[width=8cm]{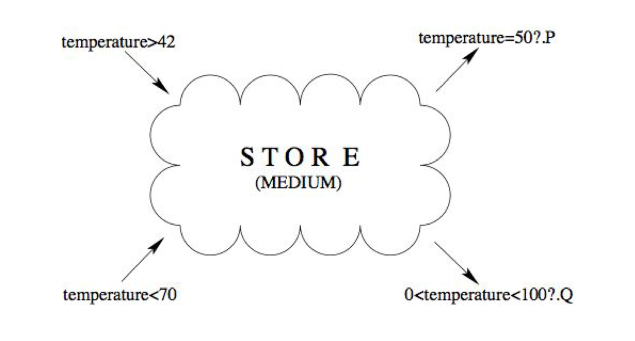}
}
\caption{Processes communicating via a common store}
\label{fig:ccp}
\end{figure}
\begin{eqnarray*}
 P, Q, \ldots & ::= &  \nul  \\
              &  |  &  \tell(c)  \\
              &  |  &  \when \ c \ \doo \ P \\
              &  |  &  P \; \Vert \; Q  \\
              &  |  &  (\local \ x; c) \ \inn \ P  \\
              &  |  &  \next \ P  \\
              &  |  &  !_{T} P  \\
              &  |  &  A(\tilde{x}) \\
\end{eqnarray*}

The null process $\nul$ does nothing. The process $\tell(c)$ adds constraint $c$ to the store within the current time. 
The process $\when \ c \ \doo \ P$ executes $P$ if  its
guard constraint $c$ is entailed by the store in the current time. Otherwise, $P$ is discarded. $P \; \Vert \; Q $ 
represents $P$ and $Q$ acting concurrently.
The process $(\local \ x; c) \ \inn \ P$  behaves like $P$, except that it declares variable $x$ private to $P$ 
(its value is hidden to other processes).
In $\next \ P$, the process $P$ will be activated in the next time unit. 
The operator $!_{T}$ is used to define infinite periodic behavior.
$!_{T} P$ represents $P \; \Vert \; \next^{T} P \; \Vert \; \next^{2T} P \; \Vert \; \cdots$ where $\next^{T} P$ is the abbreviation of
$\next \ (\next \ (\cdots (\next \ P) \cdots))$ where $\next$ is repeated $T$ times. The process $A(\tilde{x})$ is an {\it identifier} with
arity $|\tilde{x}|$. We assume that every such an identifier has a unique (recursive) definition of the form $A(\tilde{x}) \deff P$.

\subsection{Operational Semantics}
\label{sec:semantics}
The dynamics of the calculus is specified by means of two transition relations between configurations 
$\longrightarrow, \Longrightarrow \subseteq \config \times \config$
obtained by the rules in Table~\ref{tab:semantics}. 
A configuration is a pair $\langle P, d \rangle \in \proc \times \const$ where $d$ represents 
 the current store. $\config$ denote the set of all configurations with typical elements $\Gamma, \Gamma', \ldots$. 

An {\em internal transition}  $\langle P, d \rangle \longrightarrow \langle P', d' \rangle$ 
means that $P$ under the current store $d$ evolves internally
into $P'$ and produces the store $d'$ and corresponds to an operational step that take place during 
a time-unit. Rules  in upper part of Table~\ref{tab:semantics} define the  internal transitions. 
Rule (R-Tell) means that a $\tell$ process adds information (viz. a constraint) to the current store
and terminates. Rules (R-Ask) specify that the guard constraint of an ask process must be entailed by the current store
when it is triggered. Note that it is different to the usual semantics of an ask process in timed ccp-based calculi.
Our periodically time triggered scenario requires that the current store under which a process is triggered must entail its guard constraints. 
Otherwise, the process must be discarded in the current time and triggered again periodically. Rule (R-Par) specifies the concurrent
execution of multiple processes and assumes maximal parallelism since, typically in avionic systems, agents running concurrently are located
on different modules. In rule  (R-Loc), the process $(\local \ x; c) \ \inn \ P$  behaves like $P$, except that it binds the local variable
inside $P$. Rule (R-Per) states that in $!_{T} P$, the process $P$ is activated in the current time and then repeated periodically.
Finally, rule (R-Def) states that the identifier process $A(\tilde{x})$ behaves like $P$. Process $P[\tilde{v}/\tilde{x}]$ denotes  $P$
where each variable $x_i \in \tilde{x}$ inside $P$ is substituted by the value $v_i \in \tilde{v}$.

 In order to unfold the timed operator $\next$, we consider  {\em observable transitions}. 
 An observable transition   $P \ \Trans{(c,d)} \ R$, means that the process $P$ under the current 
 store $c$ evolves in {\em one time-unit} to $R$ and produces the store $d$. 
We say that $R$ is an {\em observable evolution} of $P$. Rule (R-Obs)  in lower part of 
Table~\ref{tab:semantics} defines the observable transitions. 
The transition $P \ \Trans{(c,d)} \ R$ is obtained from a finite sequence of internal transitions
$\langle P, c \rangle \longrightarrow^\ast   \langle Q, d  \rangle \not\longrightarrow \mbox{ where } F(Q)=R$ and 
$F: \proc \rightarrow \proc$, the {\em future function} is  defined as
{\small
\begin{equation*} 
  F(Q)  =  \left\{ 
 \begin{array}{ll}
 \displaystyle
  Q' & \mbox{if } Q= \next \; Q',\\[3mm]
 \displaystyle
  F(Q_1)  \; \Vert \; F(Q_2)  & \mbox{if } Q= Q_1  \; \Vert \; Q_2,\\[3mm]
 \displaystyle
  (\local \ x; c) \ \inn \ F(Q')  & \mbox{if } Q= (\local \ x; c) \ \inn \ Q',\\[3mm]
 \displaystyle
   F(Q')  & \mbox{if } Q= A(\tilde{v}) \mbox{ and } A(\tilde{v}) \deff Q',\\[3mm]
 \displaystyle
 \nul & \mbox{otherwise.} 
 \end{array}
 \right. \nonumber
 \end{equation*} 
 }
$\Gamma \not\longrightarrow$ means that there is no $\Gamma'$ such that $\Gamma \longrightarrow \Gamma'$.
%

\section{Modeling the Integrated Modular Avionics}
\label{sec:ima}

In this section, we illustrate our conceptual framework by modeling the IMA concept using the 
flight management system introduced in Section~\ref{sec:fms}. Throughout the rest of this paper, 
we shall consider the widely used Finite-Domain Constraint System $\fd[max]$~\cite{HentenryckSD94} where
$\Sigma$ is given by the constants symbols $0, 1, 2, . . . , max - 1$ 
and the relation symbols $=, \neq, <, \leq, >, \geq$.
$\Delta$ is given by the axioms in Number Theory.
%
%

\subsection{Partitions}
\label{sec:partition}

\begin{figure}[!b]
\setlength{\unitlength}{1cm}
\centerline{
\includegraphics[width=8cm]{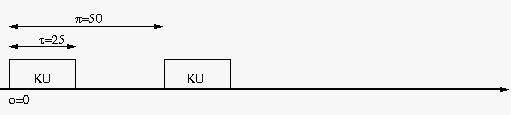}
}
\caption{FMS's Keyboard and Cursor Control Unit partition (KU)}
\label{fig:part}
\end{figure}

We assume that each partition is a black box with an input satisfying some (application dependent)
constraint $c$ and an output (viz. its result $r$) which is added to the (local) store by assigning it 
to the (local) variable $x$. Moreover, each partition (e.g. $KU$ depicted in Figure~\ref{fig:part}) is characterized 
by its offset $o$, duration $\tau$, and period $\pi$. Hence, we have the following definition:
\begin{align}
\label{eq:rtcc-partion}
&P(o, \tau, \pi) \deff \nonumber \\
& !_\pi \; (\local \; x, c_p) \; \inn \Bigl( \next^o \bigl(\when \; c \; \doo \; \next^\tau \tell(x = r) \bigr) \Bigr)
\end{align}
Intuitively, the partition $P(o, \tau, \pi)$  starts its first execution after $o$ (its offset)
time units, when the constraint $(\when \; c)$ on its input is checked. If the current store entails
$c$ then the partition adds the constraint $x = r$ to the (local) store after $\tau$ time units,
the duration of the partition. Then every $\pi$ period time, the partition is reactivated thanks 
to the $!_\pi$ operator. Note that we use $(\local \; x, c_p) \; \inn$ in the above definition only if
we do not want the current execution of the partition to overwrite the result of its previous
execution. In other words, the partition is in a queuing mode. Note that we could use 
streams~\cite{SaraswatRP91} to represent changes in the value of $x$ instead of binding $x$
locally inside $\Bigl( \next^o \bigl(\when \; c \; \doo \; \next^\tau \tell(x = r) \bigr) \Bigr)$.

\paragraph*{Example}
 Consider the $KU$ partition modeling the Keyboard and Cursor Control Unit shown in Figure~\ref{fig:part}.
 Assume that whenever $KU$ is triggered, it checks whether or not the pilot requested the waypoint information
 modelled by the boolean variable $pReq$.  If the pilot did make a request, $KU$ increment by one the waypoint ID.
 The $KU$ partition is then modelled by:
{\small
\begin{align*}
&KU(0, 25, 50) \deff \nonumber \\
& !_{50} \; \when \; (pReq = \true) \; \doo \; \next^{25} \; \tell(wpId = wpId + 1)
\end{align*}
}
since $\next^{0} \; P = P$.

\subsection{Modules}
\label{sec:module}

Now, we move onto modeling modules. We start by introducing some auxiliary functions and notations.
We define the projection functions $\sigma_i$ for any positive integer $i$, which given a vector
$\tilde{x} = (x_1, x_2, \cdots, x_n)$ returns $\sigma_i(\tilde{x}) = x_i$  ($i \leq |\tilde{x}|$).
We shall denote $s=(o, \tau, \pi)$ the scheduling\footnote{Though that only the offset time $o$ is the scheduling
parameter of a partition, by abuse of language, we will refer to the temporal parameters  $s=(o, \tau, \pi)$ 
of a partition as its scheduling parameters.} parameters of a partition and by
$\tilde{s}$ a vector of scheduling parameters ranged by their offset parameters. 

The most important requirement about the scheduling is the {\em contention-freedom} property: 
mutual-exclusion of execution times of the partitions. Let 
$$\maf(\tilde{s}) = \lcm\bigl(\{\sigma_3\bigr(\sigma_i(\tilde{s})\bigr) \; | \; 1 \leq i \leq |\tilde{s}|\}\bigr)$$ 
be the {\em least common multiple} of all partition periods of the scheduling $\tilde{s}$ which is 
commonly refereed as the {\em MAjor time Frame}. Given the scheduling vector $\tilde{s}$ 
of a module's partitions, we say that $\tilde{s}$ is contention-free, denoted $\cf(\tilde{s})$, when 
the following holds.
{\footnotesize
\begin{align}
\label{eq:cf}
&\cf(\tilde{s}) =  \true \mbox{ iff } \forall 1 \leq i \neq j \leq |\tilde{s}|, \nonumber \\
&  \forall t_i \in 
\Bigl[0 \cdot\cdot \Bigr(\frac{\maf(\tilde{s})}{\sigma_3\sigma_i(\tilde{s})} - 1\Bigl)\Bigr], 
 \forall t_j \in \Bigl[0 \cdot\cdot \Bigr(\frac{\maf(\tilde{s})}{\sigma_3\sigma_j(\tilde{s})} - 1\Bigl)\Bigr]: \nonumber \\
& \Bigr(t_i \times \sigma_3\sigma_i(\tilde{s}) + \sigma_1\sigma_i(\tilde{s}) \geq 
t_j \times \sigma_3\sigma_j(\tilde{s}) + \sigma_1\sigma_j(\tilde{s}) + \sigma_2\sigma_j(\tilde{s})\Bigl) \vee  \nonumber \\
& \Bigr(t_j \times \sigma_3\sigma_j(\tilde{s}) + \sigma_1\sigma_j(\tilde{s}) \geq 
t_i \times \sigma_3\sigma_i(\tilde{s}) + \sigma_1\sigma_i(\tilde{s}) + \sigma_2\sigma_i(\tilde{s})\Bigl) 
\end{align}
}

Now, given a scheduling vector $\tilde{s}$ of a module's partitions, the module is defined
as follows:
\begin{eqnarray}
\label{eq:rtcc-module}
M(\tilde{s}) & \deff & \when \; \cf(\tilde{s})  \; 
 \doo \; \prod_{1 \leq i \leq |\tilde{s}|} P_i\bigl(\sigma_i(\tilde{s})\bigr)
\end{eqnarray}
where each partition $P_i$  has an input satisfying some constraint $c_i$ and produces  
$r_i$ as its result. Both $c_i$ and $r_i$ are application dependent.

Informally, the module $M(\tilde{s})$ verifies that the scheduling of its partitions is 
well-defined, that is, the mutual-exclusion of partitions' window execution times is satisfied.
If the scheduling is well-defined, all the partitions are activated in the current time but due to 
the definition of partitions (see Section~\ref{sec:partition}), each of them will actually 
be triggered at its own offset time.

\paragraph*{Example}
Consider the module $M1$ of the flight management system (see Figure~\ref{fig:FMS}).
Assume that the scheduling parameters of its second partition (viz. $MFD1$ ) is 
$s_{MFD1}=(o=25, \tau=25, \pi=50)$,
then it is easy to see that the scheduling vector $\tilde{s}_{M1} = (s_{KU1},s_{MFD1})$ satisfies 
the contention-freedom property (Equation~(\ref{eq:cf})).
\begin{eqnarray*}
 M1 (\tilde{s}_{M1})  & \deff & \when \; \cf(s_{KU1},s_{MFD1})  \;  \doo \; \\
                       &      & KU1 (s_{KU1}) \; \Vert \; MFD1 (s_{MFD1}).
\end{eqnarray*}

\subsection{IMA}
\label{subsec:ima}

An \emph{IMA} system, is simply a product of multiple modules running concurrently and communicating
throughout a TTEthernet network which we will address in the following section. Given an \emph{IMA}
system of $n$ modules $\{M_1, M_2, \cdots, M_n\}$, with $\tilde{s}_i$ the scheduling vector
of $M_i$, we call {\em IMA scheduler} and denote by $ \gamma_{IMA} = \{\tilde{s}_i : 1 \leq i \leq n\}$ 
the set of the scheduling vectors of all the modules composing the system.
Table~\ref{tab:ima-schedule} shows an example of an \emph{IMA }scheduler for the flight management system.
Then, the whole \emph{IMA} system is modelled as follows.
\begin{eqnarray}
\label{eq:rtcc-ima}
 \ima(\gamma_{IMA} ) & \deff & \prod_{1 \leq i \leq |\gamma_{IMA}|} M_i(\tilde{s}_i)
\end{eqnarray}
\begin{table}[!tb]
	\centering
	\[
		\begin{array}{|l|l|l|l|l|}
		\hline
		\mbox{Partitions} & \pi & \tau & o & \mbox{Module}\\
		\hline
		\hline
		\mbox{KU1 (MFD1)} & 50 & 25 & 0 \; (25) & \mbox{M1}\\
		\hline
		\mbox{KU2 (MFD2)} & 50 & 25 & 0 \; (25) & \mbox{M2}\\
		\hline
		\mbox{FM1} & 60 & 30 & 7 & \mbox{M3}\\
		\hline
		\mbox{FM2} & 60 & 30 & 27 & \mbox{M4}\\
		\hline
		\mbox{NDB} & 100 & 20 & 77 & \mbox{M5}\\
		\hline
		\end{array}
	\]
	\caption{IMA-schedule}
	\label{tab:ima-schedule}
\end{table}

\section{Modeling TTEthernet}
\label{sec:tte}
 
 The modeling of the TTEthernet network will follow the same principle as in the previous section.
 We start by modeling tt-frames. Then we model dataflow links. Finally, in order to build the complete
 network, we piece together all datalinks if the temporal parameters of the frames satisfy all the temporal
 requirements of the TTEthernet.

\subsection{Frames}
\label{sec:frame}

According to the aerospace standard AS6802~\cite{TTE2011}, a TT frame $f_i$ on a data link $[v_k, v_l]$, denoted
$f_i^{[v_k, v_l]}$ is fully temporally specified by its {\em offset time}, {\em length} and {\em period}:
$$f_i^{[v_k, v_l]} = (f_i^{[v_k, v_l]}\cdot offset, f_i \cdot length, f_i\cdot period).$$ 
The length and the period of a frame are given a priori and remain fixed along the virtual link.
It is the task of the tt-scheduler to assign values to the frame's offset times on all dataflow links 
belonging to the frame's virtual link. 
Note that these three temporal parameters are the same that fully characterize a partition on a module
(see Section~\ref{sec:partition}). Hence, there is a perfect similarity between partitions on a module and
frames on a dataflow link. Therefore, we model a frame $f_i^{[v_k, v_l]} = (o, \tau, \pi)$ by the following
process:
\begin{align}
& F_i^{[v_k, v_l]}(o, \tau, \pi) \deff \nonumber \\
& !_\pi \; (\local \; x, c_i) \; \inn \Bigl( \next^o \bigl(\when \; c \; \doo \; \next^\tau \tell(x = r) \bigr) \Bigr)
\label{eq:rtcc-frame}
\end{align}
\noindent
where $o= f_i^{[v_k, v_l]}\cdot offset$, $\tau= f_i \cdot length$, and $\pi = f_i\cdot period$.
Again, $(\local \; x, c_i) \; \inn$ is used only if the tt-frame is transmitted under a queuing mode.

\paragraph*{Example}
Consider the waypoint ID $wpId$ produced by the $KU$ partition in the previous section. Assume that
it is transmitted along the dataflow link $[M1, SW1]$ w.r.t. the tt-scheduling parameters
$(50, 2, 10)$. Then under a sampling mode, $wpId$ along $[M1, SW1]$ is modelled as follows.
{\small
\begin{align}
& WPID1^{[M1, SW1]}(50, 2, 10) \deff \nonumber \\
& !_{10} \; \next^{50} \; \when \; (wpId1 > 0) \; \doo \; \next^2 \tell(sw11 = wpId1) \nonumber
\label{eq:rtcc-frame}
\end{align}
}
We use the guard constraint\footnote{Assuming the initial value is $wpId = 0$.} $(wpId > 0)$ 
so that the waypoint ID is transmitted only when the pilot's request is processed by $KU1$.

\subsection{Dataflow link}
\label{sec:datalink}

Exploiting the temporal characterization similarity between tt-frames and IMA partitions, 
we naturally model dataflow links the same way we modelled modules. Indeed, like
partitions on the same module, the temporal parameters of tt-frames transmitted along the same dataflow link
must satisfy the contention freedom property given by Equation~(\ref{eq:cf}). Therefore, given $\ttsched^{[v_i, v_j]}$,
the tt-scheduling vector of frames tranallowssmitted along the dataflow link $[v_i, v_j]$, 
we model the dataflow link by the following process.
\begin{eqnarray}
\label{eq:link}
L^{[v_i, v_j]}(\ttsched^{[v_i, v_j]}) & \deff & \when \; \cf(\ttsched^{[v_i, v_j]})  \; 
 \doo \nonumber \\
 & & \prod_{1 \leq i \leq |\ttsched^{[v_i, v_j]}|} F_i^{[v_i, v_j]}\bigl(\sigma_i(\ttsched^{[v_i, v_j]})\bigr)
\end{eqnarray}
%
\begin{table}[!t]
	\centering
	\[
		\begin{array}{|l|l|l|l|l|}
		\hline
		\mbox{Frames} & \pi & \tau & o & \mbox{Datalink}\\
		\hline
		\hline
		\mbox{wpId1} & 10 & 2 & 50 & \mbox{[M1,SW1]}\\
		\hline
		\mbox{wpId2} & 10 & 2 & 50 & \mbox{[M2,SW1]}\\
		\hline
		\mbox{wpId1 (wpId2)} & 10 & 2 & 55 \; (53) & \mbox{[SW1,M3]}\\
		\hline
		\mbox{wpId1 (wpId2)} & 10 & 2 & 55 \; (53) & \mbox{[SW1,M4]}\\
		\hline
		\mbox{query1}  & 30 & 3 & 40 & \mbox{[M3,SW2]}\\
		\hline
		\mbox{query2}  & 30 & 3 & 60  & \mbox{[M4,SW2]}\\
		\hline
		\mbox{query11(query2)} & 30 & 3 & 44 \; (41) & \mbox{[SW2,M5]}\\
		\hline
		\end{array}
	\]
	\caption{TT-scheduler}allows
	\label{tab:tt-schedule}
\end{table}
\paragraph*{Example}
consider the dataflow link $[SW1,M3]$ which transmits both frames $wpId1$ and $wpId2$ 
from the switch $SW1$ to the module $M1$. From the TT-scheduling parameters 
given in Table~\ref{tab:tt-schedule}, we have that
$\ttsched^{[SW1,M3]}=\bigl((55, 2, 10),(53, 2, 10)\bigr)$, which satisfies 
the contention-freedom property (\ref{eq:cf}).
\[
\begin{array}{lll}
 L^{[SW1,M3]} \Bigl((55, 2, 10),(53, 2, 10)\Bigr)  & \deff & \\
    \multicolumn{3}{l}{\; \; \when \; \cf \Bigl(\bigl((55, 2, 10),(53, 2, 10)\bigr)\Bigr) \; \doo} \\
    \multicolumn{3}{l}{\; \;  WPID1^{[SW1,M3]} (55, 2, 10) \; \Vert \; WPID2^{[SW1,M3]} (53, 2, 10).}
\end{array}
\]
\subsection{The network}
\label{sec:network}

Now that we have built each dataflow link separately, we need to piece them together in order to obtain the full
network. However, unlike the \emph{IMA} modules which operate independently from each other, dataflow links form paths and 
virtual links and hence their combination must satisfy some specific constraints. In this paper, we will illustrate
two of them: a path dependent constraint and a virtual link dependent one. For the complete list of these constraints
and their formalization, we refer to \cite{BejiHGM2014,steiner2010evaluation}.

\paragraph*{Well-formed path}
within the dataflow path of a frame the dispatch points in time on {\em two
adjacent} datalinks will be well-timed. Formally,
\begin{align}
\label{path-dependent}
&\forall p \in \dpath, \forall [v_{x},v_{y}],[v_{y},v_{z}] \in p, \nonumber \\
&(f_{i}^{[v_{y}, v_{z}]}.offset)-(f_{i}^{[v_{x}, v_{y}]}.offset) \geq max(\emph{hopdelay})
\end{align}
where $max(\emph{hopedelay})$ is an off-line configurable upper bound of the maximum latency over a single hop.

For example, assume that $max(\emph{hopedelay})=3$, then the $wpId1$ frame's path $p= [[M1,SW1],[SW1,M3]]$ 
from $M1$ to module $M3$, whose temporal parameters are given in Table~\ref{tab:tt-schedule}, 
is well-formed. We have
\begin{eqnarray*}
WPID1^{[SW1,M3]}\cdot offset &\geq & WPID1^{[M1,SW1]}\cdot offset\\
                                  && + \; max(\emph{hopedelay})
\end{eqnarray*}
\noindent
since $55 \geq 50+3$.

\paragraph*{Simultaneous relay}
an avionic functionality might require that some frames to be simultaneously dispatched 
on all outgoing dataflow links of the relaying nodes within their virtual links. Given
a virtual link $\vl$ of a frame $f_i$, the simultaneous relay requirement is satisfied
if the following holds.
\begin{align}
\label{Simultaneous-Relay}
&\forall p_{k},p_{l} \in \vl \; (k \neq l),
\forall [v_{x},v_{y}] \in p_k, \forall [v_{x},v_{z}] \in p_{l}, \nonumber \\
&(f_{i}^{[v_{x}, v_{y}]}.offset)=(f_{i}^{[v_{x}, v_{z}]}.offset)
\end{align}

For example, the switch $SW1$ dispatches simultaneously the waypoint ID frames on both $[SW1,M3]$ and 
$[SW1,M4]$ datalinks (see the tt-scheduler in Table~\ref{tab:tt-schedule}).

We are ready to build the full network. Let $\wf$ and $\sr$ denote the predicates specifying the well-formed
and the simultaneous-relay requirements respectively. Given $\ttsched_{TT}$,  the tt-scheduler of the full
network, we proceed as follows: first, we verify that $\ttsched_{TT}$ satisfies both $\wf$ and $\sr$ constraints;
then we build each datalink of the network separately; finally we piece them together thanks to the parallel 
composition operator. 
\begin{eqnarray}
  \tte(\ttsched_{TT}) &\deff & \when \; \bigl(\wf(\ttsched_{TT}) \wedge \sr(\ttsched_{TT})\bigr) \; \doo \nonumber \\ 
  &&\prod_{1 \leq i \leq |\ttsched_{TT}|} L^{[v_k, v_l]}_i\bigl(\sigma_i(\ttsched_{TT}^{[v_k, v_l]})\bigr)
\end{eqnarray}

\section{Modeling the full system}
\label{sec:avio}

\begin{figure}[!ht]
  \centering
  \includegraphics[width=0.8\linewidth]{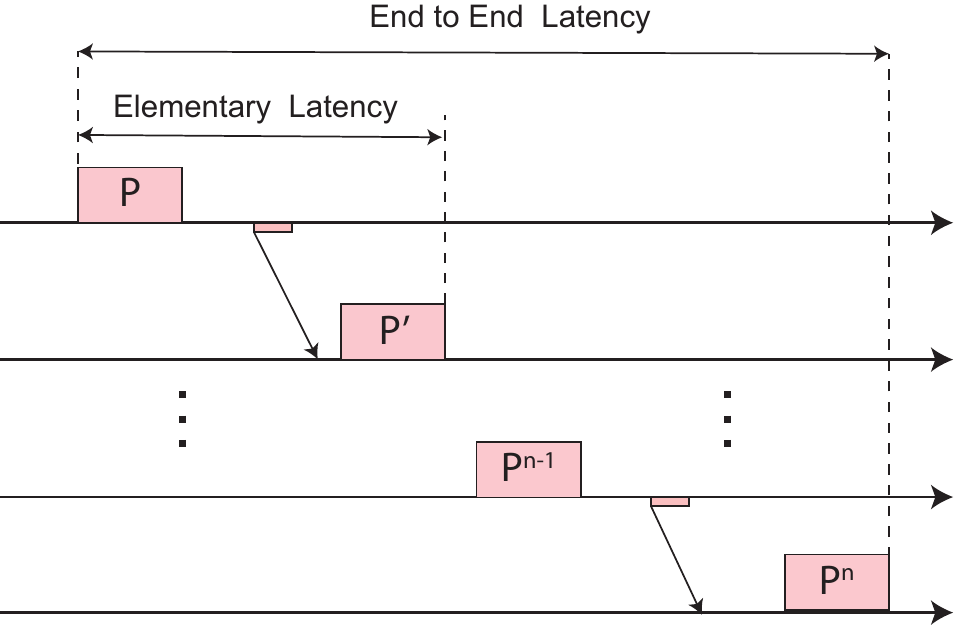} \\
  \caption{Latency}
  \label{elementaryLatency}
\end{figure}

In Section~\ref{sec:ima} and Section~\ref{sec:tte} we have built the \emph{IMA} modules and the network independently. 
In order to piece them together, the \emph{IMA} scheduler and the tt-scheduler must satisfy the 
latency constraints which ensure that some avionic functions produce their responses within some given deadlines. 
There are two types of latency: the \textit{elementary latency} and  the \textit{end-to-end latency} as depicted in
Figure~\ref{elementaryLatency}.

The elementary latency of a frame is the time duration starting from the beginning of its sending partition 
until the end of its last receiving partition. The end-to-end latency constrains the response time of 
any avionic functionality involving several communicating partitions distributed over multiple modules.
For example, we might want the flight management system to display the waypoint information within $600ms$
when a pilot makes a request. We refer the reader to \cite{BejiHGM2014} for more detail and the formalization
of the latency constraints. Let the predicate $\lt$ denotes the latency constraints, then the full system
is formalized as follows. 
\begin{eqnarray}
  \avio(\ttsched_{TT}, \tilde{\gamma}_{IMA}) & \deff \; \; \when \; \lt(\ttsched_{TT}, \tilde{\gamma}_{IMA}) \; \doo \nonumber \\
	& \ima(\tilde{\gamma}_{IMA}) \; \Vert \; \tte(\ttsched_{TT}).
\end{eqnarray}

\section{Concluding Remarks}
\label{sec:concl}

This paper presents a new simple and elegant ccp-based calculus for the analysis of real-time systems tailored for
the time-triggered processes. We show the applicability of the calculus by an elegant modeling of the concepts related 
to the \emph{IMA} and TTEthernet architectures. We also illustrate these concepts by a subsystem of the well-known 
flight management system. 

In this exposition, however, we have considered only time-triggered traffic whilst the
TTEthernet enables time-triggered and event-triggered communication, as well as integrated 
time-triggered/event-triggered communication on the same physical network. Our future work
includes the integration of event-triggered communications as well as the study of the impact
of time-triggered traffic over the latency of event-triggered traffic.

We also plan to develop a set of reasoning techniques tailored for the verification of the requirements
of avionic systems. These requirements include the \textit{non-interference} between any low level critical
entity and a higher level critical entity. For example, the complete absence of any causal failure propagation 
from low level entities to high level ones. Another interesting property is the \textit{redundancy} which is very common
in avionic systems. We plan to develop reasoning techniques for ensuring that, from an observational point of view, 
a redundant system is ``equivalent'' to its non-redundant counterpart, for example in the absence of failures.  

Finally, we plan to develop a general methodology and an associated tool for translating AADL~\cite{AADL2012}
(Architecture Analysis and Design Language) and Annexes specification (e.g.~\cite{TiyamEGHM2014}) into the $\rtcc$ language to allow 
a comprehensive analysis for avionic systems specified in this aerospace standard for model-based 
specification of complex real-time embedded systems.
\subsubsection*{Acknowledgments} This research is supported in part by  the Collaborative Research and Development (CRD) grant No.  435325-12 jointly funded by the  Consortium for Research and Innovation in Aerospace in Qu\'ebec (CRIAQ) and the Natural Sciences and Engineering Research Council of Canada (NSERC) as part of project {\em Verification of Integration of Time-Triggered Avionic Systems} (VerITTAS).
\bibliographystyle{abbrv}

\end{document}